\documentclass[preprints,article,submit,submit,pdftex,moreauthors]{Definitions/mdpi}

\usepackage{amsthm}
\usepackage{amsmath}
\usepackage{graphicx}
\usepackage{booktabs}
\usepackage{mwe}
\usepackage{svg}
\usepackage{float}
\usepackage{multirow}
\usepackage{multicol}
\usepackage{relsize}
\usepackage{geometry}
\usepackage{amssymb}
\usepackage{amsfonts}
\usepackage{subcaption}
\usepackage{numprint}
\npdecimalsign{.}
\nprounddigits{2}
\firstpage{1} 
\pubvolume{1}
\issuenum{1}
\articlenumber{0}
\pubyear{2023}
\copyrightyear{2023}
\datereceived{ } 
\daterevised{ } 
\dateaccepted{ } 
\datepublished{ } 
\hreflink{https://doi.org/} 

\renewcommand{\linenumbers}{}

\Title{Understanding Self-Supervised Learning of Speech Representation via Invariance and Redundancy Reduction}

\TitleCitation{Understanding Self-Supervised Learning of Speech Representation via Invariance and Redundancy Reduction}

\Author{Yusuf Brima $^{\star,1,2}$, Ulf Krumnack $^{1}$, Simone Pika $^{2}$, and Gunther Heidemann $^{1}$}

\AuthorNames{Yusuf Brima, Ulf Krumnack, Simone Pika and Gunther Heidemann}

\AuthorCitation{Brima, Y.; Krumnack, U.; Pika, S., Heidemann, G.}

\address{%
$^{1}$ \quad Computer Vision, Institute of Cognitive Science, Osnabrueck University\\
$^{2}$ \quad Comparative BioCognition, Institute of Cognitive Science, Osnabrueck University}

\corres{Correspondence: ybrima@uos.de; }

\abstract{
Self-supervised learning (SSL) has emerged as a promising paradigm for learning flexible speech representations from unlabeled data. By designing \textit{pretext tasks} that exploit statistical regularities, SSL models can capture \textit{useful} representations that are \textit{transferable to downstream tasks}. This study provides an empirical analysis of Barlow Twins (BT), an SSL technique inspired by theories of redundancy reduction in human perception. On downstream tasks, BT representations accelerated learning and transferred across domains. However, limitations exist in disentangling key explanatory factors, with redundancy reduction and invariance alone insufficient for factorization of learned latents into \textit{modular}, \textit{compact}, and \textit{informative} codes. Our ablations study isolated gains from invariance constraints, but the gains were context-dependent. Overall, this work substantiates the potential of Barlow Twins for sample-efficient speech encoding. However, challenges remain in achieving fully hierarchical representations. The analysis methodology and insights pave a path for extensions incorporating further inductive priors and perceptual principles to further enhance the BT self-supervision framework.
}


\begin{document}
\section{Introduction}
Speech processing holds a pivotal role in diverse applications, spanning speaker identification, diarization, spoken language understanding, speaker segmentation, voice assistants, etc. \cite{togneri2011overview,tirumala2016review,lukic2016speaker,trong2016deep,adaloglou2021comprehensive}. The extraction of linguistic and para-linguistic features from speech data is essential for ensuring accurate and robust performance within these application domains. Despite the effectiveness of conventional supervised learning methods\cite{tirumala2016review,bhangale2021review}, their heavy reliance on labels as supervisory signals poses challenges due to the scarcity and expense associated with obtaining such labels~\cite{mohamed2022self,kemp1999unsupervised,lamel2002lightly}.

Self-supervised learning (SSL) has emerged as a paradigm for learning flexible representations from \textit{unlabeled} data by exploiting inherent statistical regularities as supervisory signals. A core tenet of SSL is designing \textit{pretext tasks} to train deep learning models to capture intrinsic statistical structures within inputs  without the need for human labeling. For speech, abundant redundancies exist within audio regarding linguistic content, speaker characteristics, emotions, etc. SSL leverages these ubiquitous patterns in speech through extensive use of data augmentation and context-based predictive pretext tasks. These include predicting masked time-frequency spectrogram components from neighboring regions or contrastive learning objectives judging different (distorted) versions of the same underlying utterance as identical~\cite{oord2018representation,chung2019unsupervised}. Such techniques enable models to focus representations on speaker and/or language information while discarding nuisance variations such as background noise.

A recently proposed cognitive neuroscience-inspired framework builds upon progress in SSL for speech by aligning with principles of redundancy reduction characterized by Horace Barlow~\cite{barlow2001redundancy}. Specifically, Barlow Twins (BT) adopts a \textit{joint embedding architecture} (JEA) trained to produce consistent encoder representations between differently augmented views of the same speech input~\cite{zbontar2021barlow}. This, in the context of the speech, aims to emulate auditory sensory perception efficacy in amplifying speaker-related cues while suppressing irrelevant variation. The integration of core redundancy minimization concepts and the general SSL paradigm offer promise in improving the sample efficiency, flexibility, and biological plausibility of self-supervised speech encoding techniques to build \textit{robust cognitive schema of auditory representations}.

However, the utility of this framework in achieving \textit{distributed}, \textit{disentangled}, and \textit{invariant} representations remains underexplored. Therefore, this paper undertakes an empirical analysis on three fronts to address open questions:

\begin{itemize}
    \item \textbf{Downstream Task Efficacy}: Quantitative effectiveness on a select speech processing task.
    \item \textbf{Disentanglement Analysis}: A quantitative assessment of representation decoupling quality and its axis alignment with ground-truth explanatory factors.
    \item \textbf{Objective Variants}: Ablations examine impact of training components.
\end{itemize}

This investigation systematically examines the BT framework, focusing specifically on the representational quality and performance attributable to its redundancy reduction principles. Our study, centered on analytical evaluation rather than competitive benchmarking, provides novel evidence about the framework's suitability for learning useful speech representations. The aim is to substantiate the utility of BT while identifying potential advancements for sample-efficient speech encoding models. Although an exhaustive benchmarking of various methods falls beyond this study's scope, our contribution lies in the rigorous assessment of Barlow Twins. We evaluate this framework as a \textit{simple} yet effective approach, particularly in the realms of invariance and redundancy reduction, crucial for learning useful speech representations.

We structure the subsequent sections as follows. In Section~\ref{mats}, the Materials and Methods detail our proposed self-supervised speech representation learning framework. Section~\ref{res} benchmarks performance across diverse speech tasks and datasets, quantifying emerging representation quality through \textit{disentanglement} analyses and ablation studies of loss objective variants. Section~\ref{dis} analyzes result outcomes. Finally, Section~\ref{con} encapsulates key contributions, synthesizes insights derived from our study, and provides a succinct summary, offering a conclusive wrap-up to the paper.
\subsection{Related Works}
Self-supervised learning (SSL) methods have gained traction in speech processing for their ability to learn representations without manual annotations~\cite{liu2021self,liu2022audio,khosla2020supervised,chen2020simple,grill2020bootstrap,caron2020unsupervised,oord2018representation,koch2015siamese}. In this context, the complexities of existing SSL techniques, such as wav2vec 2.0~\cite{baevski2020wav2vec} and HuBERT~\cite{hsu2021hubert}, often involve specialized negative sampling, stop gradients, and \textit{intricate training recipes}. These complexities, while contributing to the effectiveness of these methods, can also pose challenges to their flexibility and adaptability.

Wav2vec 2.0 and HuBERT represent state-of-the-art SSL techniques in speech processing. Wav2vec 2.0 employs a contrastive learning approach, where the model learns to distinguish between positive and negative samples by maximizing agreement between positive pairs and minimizing it between negative pairs~\cite{baevski2020wav2vec}. This requires careful handling of negative samples and intricate training recipes to ensure convergence and effectiveness.

HuBERT, on the other hand, focuses on a masked language modeling approach combined with contrastive learning, leveraging hierarchical structures for representation learning~\cite{hsu2021hubert}. The model involves complex strategies such as predictive masking of hidden units and k-means clustering to enhance the quality of speech embeddings. These methods, while successful, introduce challenges related to the need for specialized negative sampling and the delicate balance required during training.

In contrast, the Barlow Twins (BT) framework offers a conceptually \textit{simpler} SSL approach, relying only on data augmentation (\textit{multi-view creation)}, and a redundancy reduction and invariance objective~\cite{zbontar2021barlow}. This simplicity is achieved through maximizing cross-correlation between augmented views of inputs while minimizing cross-sample cross-correlation.

Motivated by this revelation of simplicity versus complexity trade-offs, this study seeks not to outperform state-of-the-art, but rather to conduct an extensive empirical analysis quantifying the utility of adopting the BT framework for speech representation learning. Through an evaluation of diverse downstream speech tasks and datasets, we center our investigation on assessing learned representation quality along pertinent axes of generalization, disentanglement, and factorial representation of key speech factors. 
\section{Materials and Methods}
\label{mats}
\subsection{Learning Framework}
\begin{figure}[th!]
\centering
\includegraphics[width=1.0\textwidth]{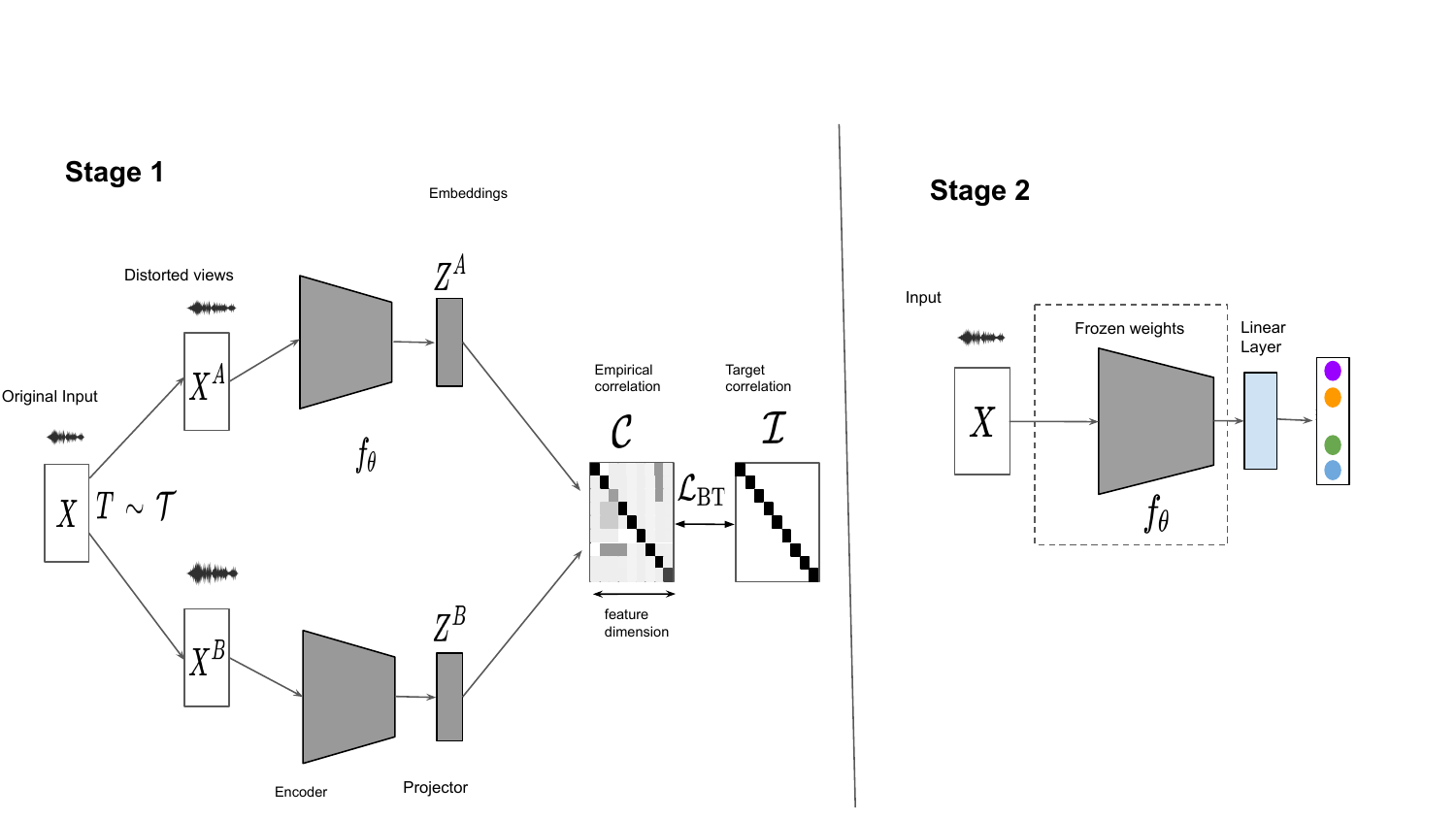} 
\caption{The BT framework for learning invariant speech representations. \textbf{Stage 1:} An encoder $f_\theta$ process augmented views $X^A$ and $X^B$ of the same speech input $X$ and project them into a shared latent space. The BT loss (Equation~\ref{eq:BT}) enforces redundancy reduction between latents from different samples while maximizing correlation for positive pairs (two views of the same sample). This causes the encoders to produce invariant representations capturing speaker identity while reducing sensitivity to augmentations. \textbf{Stage 2:} The learned latent representations $Z^A$ and $Z^B$ can then be used for downstream speech processing tasks to evaluate the model's generalization capability.}
\label{fig:Barlow_Twins}
\end{figure}

The Barlow Twins (BT) framework, as depicted in Figure~\ref{fig:Barlow_Twins}, employs a \textit{joint embedding architecture} (JEA) to learn invariant representations. Specifically, it uses an encoder network $f_{\theta}$ to project augmented views of speech within a mini-batch - denoted $X^A$ and $X^B$ - into a \textit{shared latent space}, producing latent representations $Z^A$ and $Z^B$ respectively (see Figure \ref{fig:Barlow_Twins}). The key idea is that differently, augmented views of the same underlying speech sample should have similar latent variables, while views from different samples should be de-correlated in the latent space. 

This is formalized through the two-component optimization objective in equation~\eqref{eq:BT}. First, the cross-correlation matrix $C_{ij}$ in equation~\eqref{eq:correlation} between the latent variables $z_i^A$ and $z_j^B$ for a positive pair (i.e. two augmented views of the same sample) should be close to $1$. Second, the redundancy term enforces de-correlation between latent variables from different samples. This has the combined effect of making $Z^A$ and $Z^B$ invariant for positive pairs, while also reducing redundancy across the mini-batch. Training the encoder $f_\theta$ with this learning objective thus produces a representation space with useful properties for downstream tasks.

\begin{equation}
\mathcal{L}(C;\lambda) \triangleq \underbrace{\sum_i\left(1-C_{i i}\right)^2}_{\text {invariance }}+\lambda\underbrace{\sum_i \sum_{j \neq i}\left(C_{i j}\right)^2}_{\text {redundancy reduction }}.
\label{eq:BT}
\end{equation}
\begin{equation}
\mathcal{C}_{i j} \triangleq \frac{\sum_b z_{b, i}^A z_{b, j}^B}{\sqrt{\sum_b\left(z_{b, i}^A\right)^2} \sqrt{\sum_b\left(z_{b, j}^B\right)^2}},
\label{eq:correlation}
\end{equation}

\begin{figure}[!ht]
    \centering
    \includegraphics[width=0.45\textwidth]{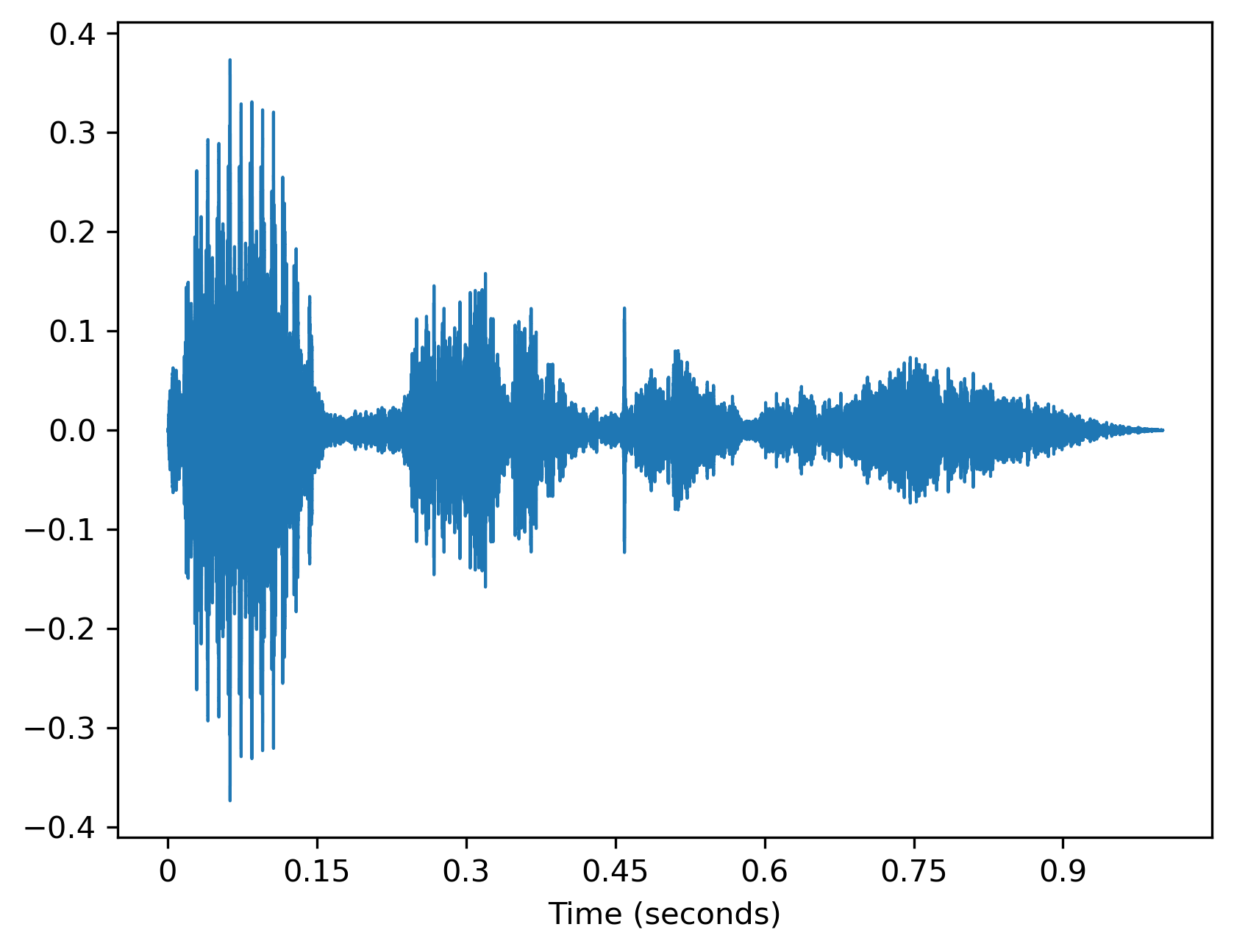}
    \includegraphics[width=0.45\textwidth]{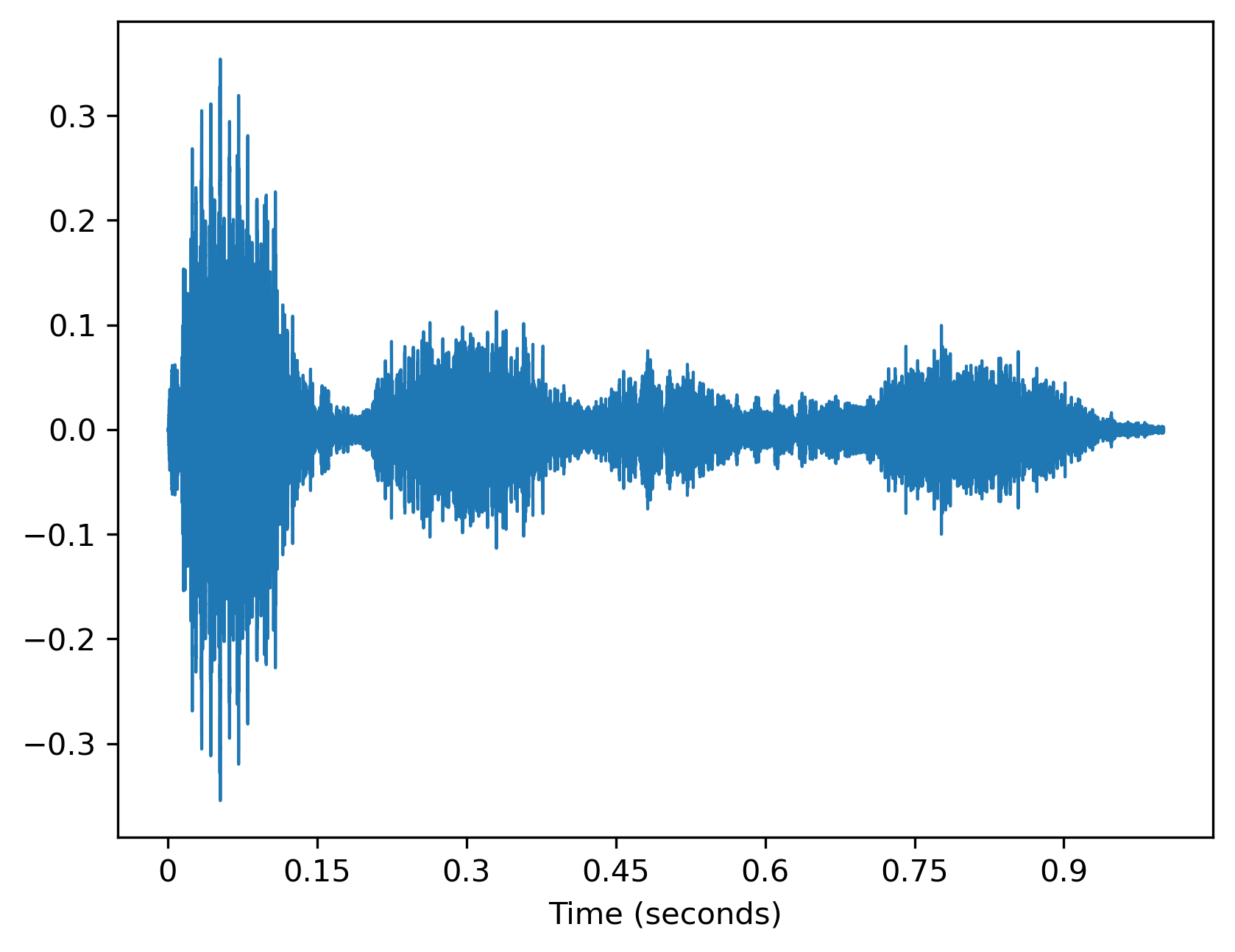}
    \includegraphics[width=0.45\textwidth]{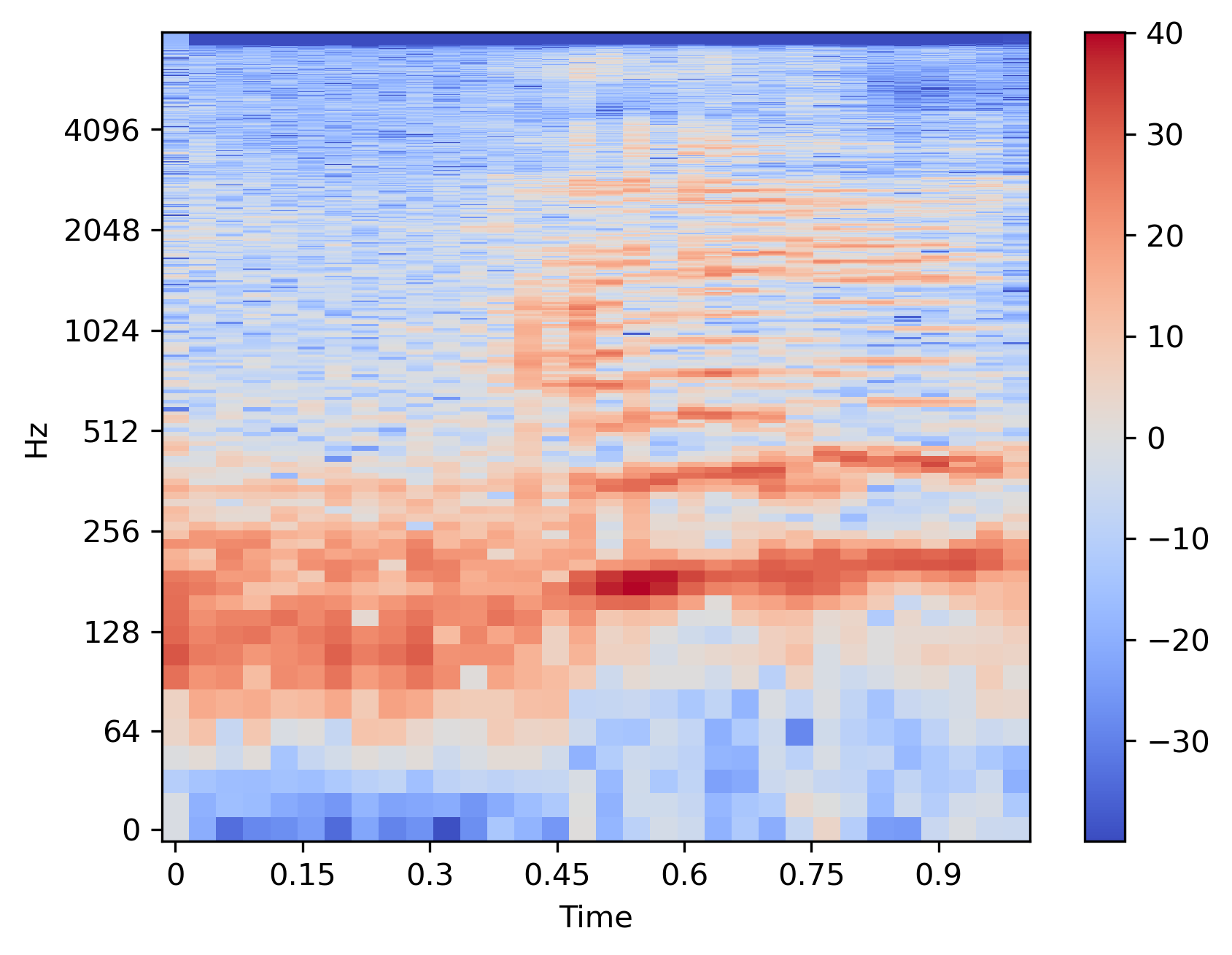}
    \includegraphics[width=0.45\textwidth]{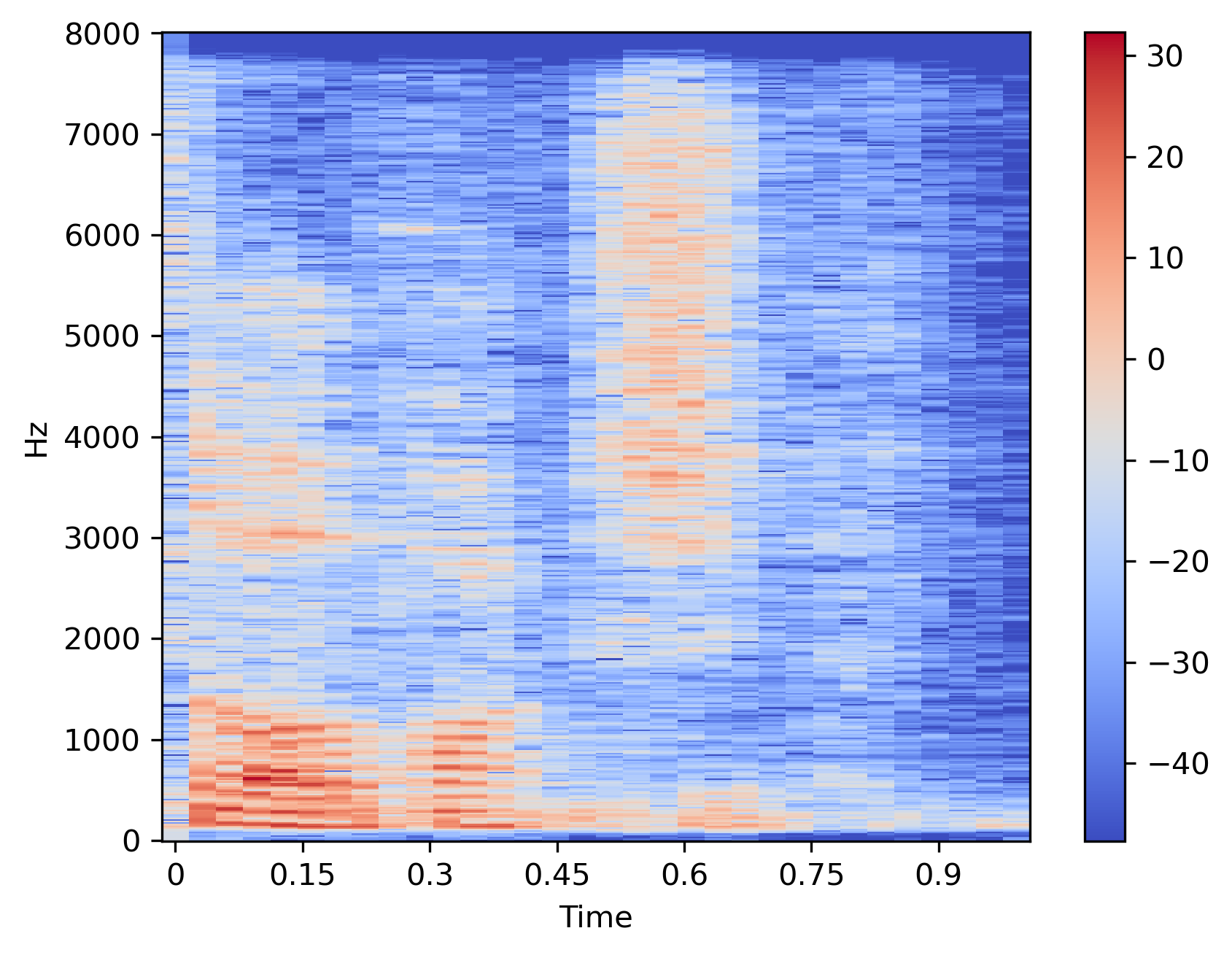}
    \caption{\textbf{(Left column)} View 1 provides a dual representation, featuring the time-domain signal (top row) and its corresponding time-frequency spectrogram (second row), both derived from the first perturbed version of the original audio signal. \textbf{(Right column)} View 2 presents a similar pair of representations. The higher harmonic partials present in the first view are not visibly present in the second view, however, the underlying information content remains invariant.}
    \label{fig:view_creation}
\end{figure}

In Figure~\ref{fig:view_creation}, we illustrate the creation of two views crucial to our SSL approach. The left column showcases View 1, offering both the time-domain representation (top row) and the corresponding time-frequency spectrogram (second row), both derived from the first perturbed version of the original audio signal. On the right column, View 2 mirrors this representation, providing a parallel set of time-domain and spectrogram views. These views, capturing variations within the input signal, form the foundation for our SSL framework, enabling the model to glean  invariant information while attenuating irrelevant variations.
\subsection{Datasets}
We utilize a diverse collection of speech datasets, summarized in Table~\ref{tab:datasets}, to train representation models (upstream) and evaluate downstream tasks. For upstream representation learning, we leverage \textit{VoxCeleb-1}, \textit{LibriSpeech-100}, and \textit{LibriSpeech-360}, which provide wide coverage of speakers and speech. VoxCeleb-1 contains over 100,000 utterances from 1,211 celebrities, while LibriSpeech-100 and LibriSpeech-360 consist of audio-book excerpt readings from 128 and 921 speakers respectively. This diversity of training data is crucial for learning robust and generalizable speech representations.

We assess the learned representations on downstream tasks utilizing the \textit{Google Speech Commands}, \textit{Emotional Speech Dataset (ESD)}, and \textit{World Leaders at the US Congress (WLUC)} datasets. Speech Commands provides a collection of spoken words for keyword spotting, ESD has emotional speech samples, and WLUC consists of worldwide leader speeches for speaker and gender identification. Performance on these downstream tasks indicates how informative and transferable the upstream representations are for speech-processing objectives.

\begin{table}[!ht]
\centering
\caption{Summary of upstream and downstream datasets. "Upstream" tasks refer to self-supervised training in our case optimizing for the BT learning objective of redundancy reduction and invariance of the multi-view representations, while "Downstream" tasks include keyword spotting, emotional tone recognition, speaker identification, and gender recognition.}
\label{tab:datasets}
\begin{tabular}{llcccl} 
\toprule
 Source & Dataset Name & \# Samples & \# Classes & Duration (hrs) & Usage \\
\midrule
~\cite{nagrani2020voxceleb} & VoxCeleb-1 & 148,642 & 1,211 & 340.39
 &Upstream \\
\cite{panayotov2015librispeech} & LibriSpeech-100 & 14,385 & 128 & 100 & Upstream\\  
~\cite{panayotov2015librispeech} & LibriSpeech-360 & 104,935 & 921 & 360 & Upstream\\
~\cite{warden2018speech} & Speech Commands & 7,985 & 2 & 2.18 & Downstream\\
~\cite{zhou2021emotional} & ESD & 7000 & 2  & 5.52 & Downstream \\
~\cite{wluc} & WLUC & 7,500 & 5 & 2.05 &  Downstream\\
\bottomrule
\end{tabular}
\end{table}
By learning representations on diverse upstream datasets and testing generalization capability through varied downstream tasks, we comprehensively evaluate the models' capabilities. The multi-dataset, multi-task framework provides a rigorous methodology for representation learning and evaluation in speech processing.
\subsection{Experimental Setup}
\label{sec:experiment}
In our experimental setup, we followed established practices for SSL and utilized a ResNet-50 backbone for pre-training, as proposed in the original paper and consistent with other SSL frameworks. This choice of backbone architecture is well-established in the literature and provides a robust foundation for learning representations from the audio datasets outlined in Table~\ref{tab:datasets}. The pre-training process involved 50 epochs for each upstream model, with a mini-batch size of $n=64$ due to computational constraints and a latent dimensionality of $m=2028$, ensuring a comprehensive exploration of the feature space. Additionally, our audio pre-processing, including standardized sampling rates and the generation of log-scaled spectrograms, laid the groundwork for effective model training and subsequent evaluation.

To facilitate optimal learning, we applied a consistent pre-processing pipeline to all audio samples. Initially, we standardized the sampling rate of the samples to 16 kHz. Subsequently, each audio segment underwent partitioning into contiguous 1-second intervals, ensuring uniform input lengths for subsequent processing.

A pivotal aspect of the feature extraction process involved the generation of log-scaled spectrograms. By employing a window size of 64 milliseconds with a 32-millisecond hop size, we captured 513 mel-frequency bins spanning the audible frequency range of 0 to 8 kHz. The resulting spectrograms, denoted as $X\in \mathbb{R}^{513\times 126}$, encapsulated both frequency and temporal information. These spectrograms formed the foundation of our neural network architecture, serving as input tensors $X_B\in \mathbb{R}^{n\times1\times513\times 126}$, where $n$ represents the mini-batch size. This audio preprocessing ensured a standardized and informative representation, crucial for effective model training and subsequent evaluation.
\section{Results}
\label{res}
\subsection{Effect of Upstream and Downstream Dataset Sizes}
Our results in Table~\ref{tab:perf_speakersss} demonstrate that downstream task performance generally improves with more in-domain data, as evidenced by the increasing accuracy with larger dataset fractions. However, we achieve substantial gains even with very small downstream sets (5-10\%) by transferring self-supervised upstream representations, validated via linear evaluation. This showcases the transferability of learned features without extensive manual annotations.

Intriguingly, we find that LibriSpeech-100, the smallest upstream corpus, drives the strongest downstream gains - achieving over 80\% on speaker and gender recognition with just 50\% target data. More notably, with full downstream sets, it exceeds the larger upstream datasets on all 4 tasks. This reveals that rather than sheer dataset size, quality is more crucial for representation generalization - aspects at which LibriSpeech-100 excels due to expert voice actors and minimal noise.
\begin{table}[h]
\centering
\caption{Top-1 test performance evaluation across 4 downstream tasks: Speaker Recognition (SR), Gender Recognition (GR), Keyword Spotting (KWS), and Emotional Tone Recognition (ER) utilizing varied fractions of these respective downstream datasets.}
\label{tab:perf_speakersss}
\begin{tabular}{llcccc}
\toprule
\toprule
& Fraction (\%) & Supervised & LibriSpeech-100 & LibriSpeech-360 & VoxCeleb1 \\ \hline
\multirow{4}{*}{\rotatebox[origin=c]{45}{SR}}  &     5             & 34.21    & 39.47           & 28.95           & 36.84     \\
    & 10            & 54.67    & 64.00           & 54.67           & 48.00     \\
   &  50            & 75.20    & 83.73           & 77.60           & 68.00     \\
   &  100           & 84.53    & 84.93           & 81.20           & 75.07     \\ 
\bottomrule
\multirow{4}{*}{\rotatebox[origin=c]{45}{GR}}  &     5             & 66.67    & 70.00           & 63.33           & 66.67     \\
   &  10            & 75.00    & 71.67           & 75.00           & 68.33     \\
  &   50            & 79.67    & 88.67           & 87.00           & 78.33     \\
  &   100           & 62.67    & 90.17           & 88.67           & 84.67     \\ 
  \bottomrule
  \multirow{4}{*}{\rotatebox[origin=c]{45}{KWS}}  &     5             & 47.50    & 52.50           & 62.50           & 45.00     \\
   &  10            & 51.25    & 52.50           & 50.00           & 50.00     \\
   &  50            & 50.76    & 55.33           & 52.28           & 47.72     \\
   &  100           & 50.32    & 75.03           & 60.08           & 53.99     \\ 
   \bottomrule
   \multirow{4}{*}{\rotatebox[origin=c]{45}{ER}}  &  5 &       48.57 &           54.29 &           51.43 &      68.57 \\
        & 10 &       61.43 &           47.14 &           44.29 &      50.00 \\
        &   50 &       46.57 &           46.29 &           50.86 &      48.86 \\
        &  100 &       83.00 &           59.00 &           51.00 &      46.29 \\
    \bottomrule
    \bottomrule
\end{tabular}
\end{table}

We show that smaller upstream datasets, while limited in volume, can unlock substantial transfer potential if the data exhibits diversity, quality, and relevance to target domains. Specifically, LibriSpeech-100, despite its modest size, drives the strongest performance owing to its inclusion of varied professional speakers and minimal artifacts. This suggests curation may supersede the raw dataset scale.

However, while transferred features accelerate downstream learning, sufficient in-domain supervision remains indispensable for maximizing absolute performance. This is evidenced by accuracy gaps with 100\% training data between self-supervised and supervised paradigms. Therefore, effectively pre-trained representations complement, rather than replace target task annotations.

Additionally, we find task complexity and similarity across domains modulate the transferability of representations. Simpler objectives like speaker recognition mature faster with less task-specific data. But complex tasks, like emotion recognition, necessitate more in-domain data. Likewise, the affinity between pre-training and targets boosts feature usability – VoxCeleb-1 specializes in speaker cues.

Thus, high-quality, diverse self-supervised pre-training can unlock substantial value from modest downstream supervision, but task complexity, dataset relevance, domain similarity, and in-domain data size interact to determine performance gains. Carefully navigating these factors is key to optimizing representation transfer from upstream tasks to domain-specific problems.
\subsection{Can enforcing redundancy reduction and invariance result in disentanglement?}
To explore the disentanglement of latent variables in the learned representations in upstream models, we employ various disentanglement metrics including Mutual Information Gap (MIG)~\cite{chen2018isolating}, Joint Entropy Minus Mutual Information Gap (JEMMIG)~\cite{do2019theory}, Disentanglement, Completeness, and Informativeness MIG (DCIMIG)~\cite{sepliarskaia2019not}, Attribute Predictability Score (SAP)~\cite{kumar2017variational} and Modularity Score~\cite{ridgeway2018learningfstatistic}. By focusing on factors such as accent, identity, and gender, we aim to quantify and evaluate the degree to which our models disentangle these specific attributes from the overall representation. These metrics provide valuable insights into the \textit{modularity}, \textit{compactness}, and \textit{informativeness} of trained BT models, shedding light on the nuanced aspects of the learned latent space.

\begin{figure}[!h]
    \centering
     \includegraphics[width=0.45\textwidth]{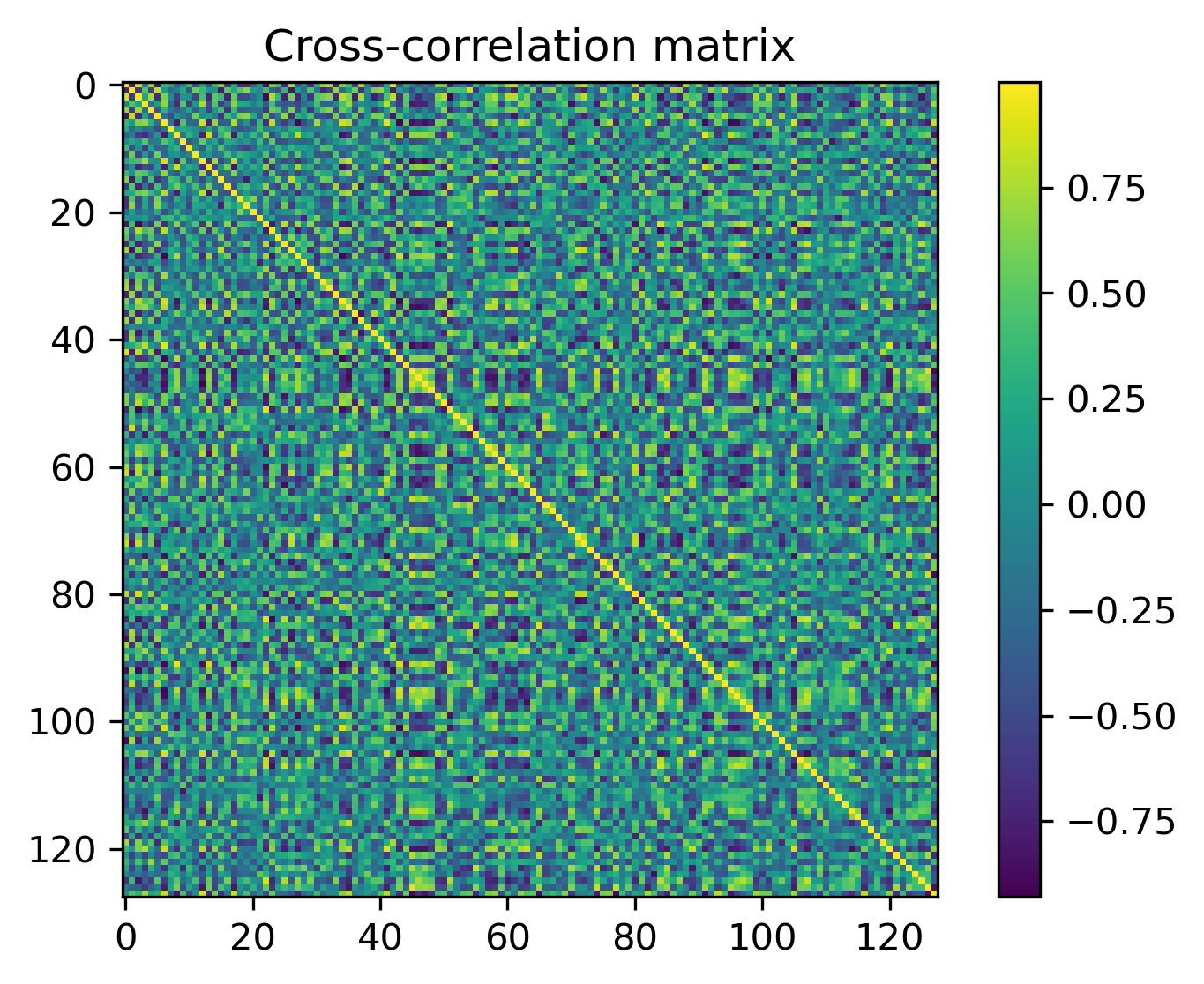}
      \includegraphics[width=0.45\textwidth]{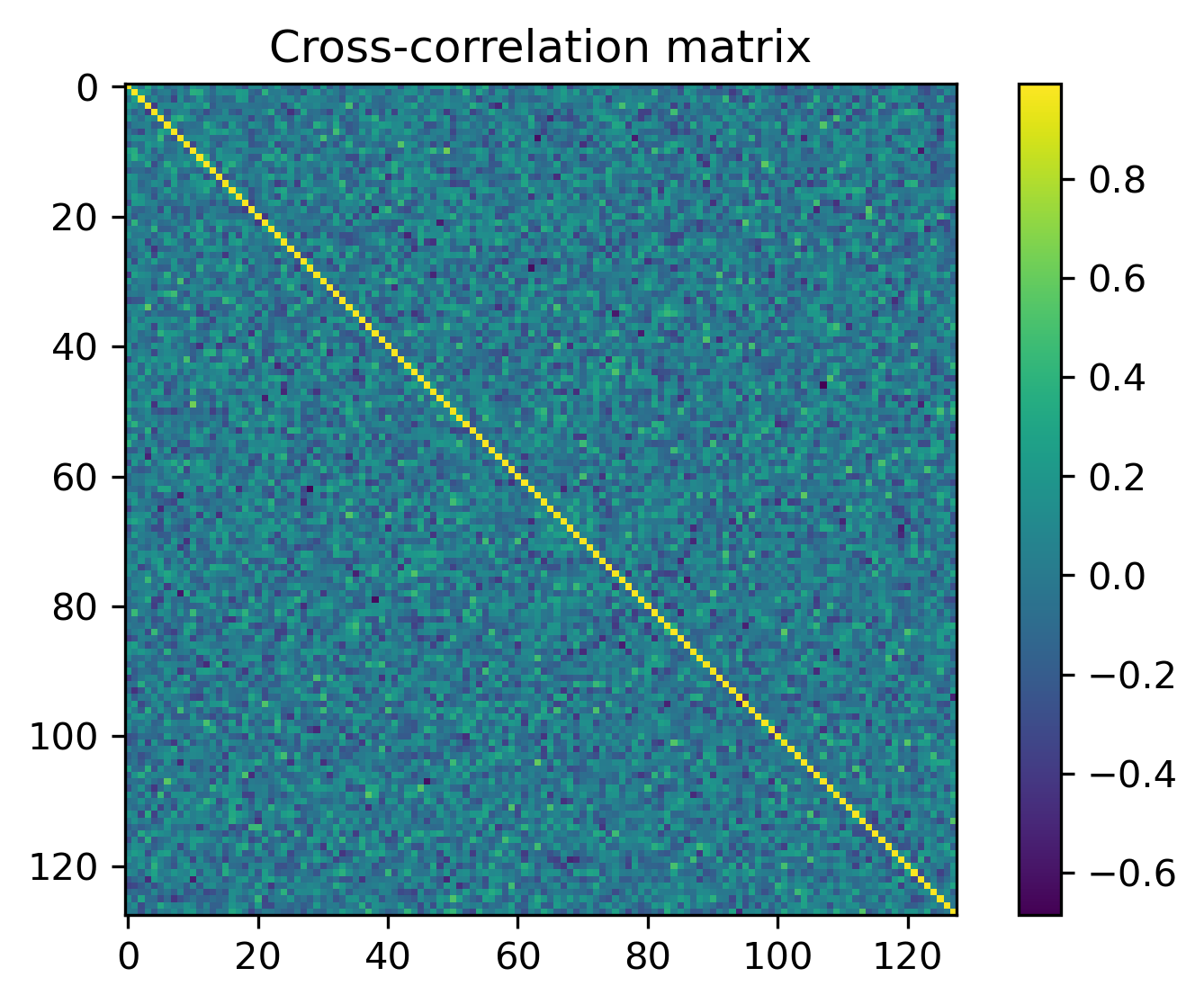}
    \caption{ Represent the empirical cross-correlation matrices, contrasting the untrained state \textbf{(left)} with the trained state \textbf{(right)} within the BT framework. These matrices visually represent the relationships between different views of the same speech input for the current mini-batch. The comparison allows us to observe the transformation in cross-correlation patterns following the self-supervised learning process, highlighting the model's ability to capture invariance (higher correlation of diagonal elements of the trained network's matrix) and de-correlation of off-diagonal elements.}
    \label{fig:emperical_cross_correlation}
\end{figure}

To visually assess the nature of learned representations, we have shown Figure~\ref{fig:emperical_cross_correlation} comparing representations of both randomly initialized and trained Barlow Twins networks. For the trained network, we observe a nearly perfect correlation along the diagonal of the cross-correlation matrix, indicating invariance between augmented views of the same input speech sample. Additionally, the off-diagonal elements are pushed closer to zero, demonstrating redundancy reduction between latents from different samples. By contrast, the untrained network shows no clear regularity. Employing such visualization techniques provides valuable insights into the structure of the learned representation space. Our analysis verifies that the BT objective successfully enforces invariance and redundancy reduction between two augmented views of the speech input. This is a crucial step in quantifying the model's capacity to capture intricate information in speech data while attenuating nuisance variation.

Analyzing the suite of disentanglement metrics in Table~\ref{tab:disentanglement-metrics}, we assess if simply enforcing redundancy reduction and invariance through the core BT learning objective can factorize learned representations along explanatory attributes without further constraints.

The consistently low scores across crucial metrics like MIG (0.020 max) and SAP (0.037 max) indicate that this training alone de-correlates but does not fully factorize key explanatory factors. While higher modularity scores (0.696 max) confirm the clustering of semantic information, specificity along individual latent dimensions remains insufficient. This highlights the need to couple complementary techniques that impose stricter decomposition for realizing fully compact and decoupled representations.

\begin{table}[h]
    \centering
    \caption{Disentanglement metrics with standard deviation for BT models over 50 evaluation runs each. Values are presented as mean $\pm$ standard deviation. BT-$n$ denotes models with a latent dimensionality of $n$. BT-LS-100 and BT-LS-360 indicate models trained on LibriSpeech-100 and LibriSpeech-360 datasets respectively. BT-VC-1 represents the model trained on the VoxCeleb-1 dataset. This table summarizes disentanglement quantification for Barlow Twins models trained on various datasets and with differing latent capacities to elucidate the impact of these factors.}
    \begin{tabular}{lcccccc}
        \toprule
        \multicolumn{1}{c}{} & \multicolumn{2}{c}{\underline{Compactness}} & \multicolumn{2}{c}{\underline{Holistic}} & \multicolumn{1}{c}{\underline{Modularity}} \\
        Model & MIG & SAP & DCIMIG & JEMMIG & Mod.~Score \\
        \midrule
        BT-16  & $0.004 \pm 0.0022$ & $0.033 \pm 0.0016$ & $0.013 \pm 0.0004$ & $0.191 \pm 0.0120$ & $0.659 \pm 0.0043$\\
        BT-32  & $0.004 \pm 0.0011$ & $0.010 \pm 0.0020$ & $0.013 \pm 0.0004$ & $0.212 \pm 0.0154$ & $0.707 \pm 0.0040$\\
        BT-64  & $0.008 \pm 0.0012$ & $0.007 \pm 0.0015$ & $0.012 \pm 0.0003$ & $0.182 \pm 0.0089$ & $0.652 \pm 0.0031$\\
        BT-128  & $0.020 \pm 0.0019$ & $0.019 \pm 0.0020$ & $0.010 \pm 0.0004$ & $0.231 \pm 0.0047$ & $0.664 \pm 0.0029$\\
        BT-512  & $0.003 \pm 0.0019$ & $0.003 \pm 0.0012$ & $0.010 \pm 0.0004$ & $0.266 \pm 0.0252$ & $0.675 \pm 0.0022$\\
        BT-1024  & $0.007 \pm 0.0010$ & $0.025 \pm 0.0018$ & $0.016 \pm 0.0004$ & $0.238 \pm 0.0175$ & $0.681 \pm 0.0021$\\
        BT-2048  & $0.005 \pm 0.0015$ & $0.014 \pm 0.0020$ & $0.012 \pm 0.0005$ & $0.294 \pm 0.0082$ & $0.691 \pm 0.0018$\\
        \midrule
        BT-LS-100 & $0.007 \pm 0.0012$ & $0.016 \pm 0.0024$ & $0.013 \pm 0.0005$ & $0.141 \pm 0.0091$ & $0.685 \pm 0.0021$\\
        BT-LS-360 & $0.006 \pm 0.0012$ & $0.037 \pm 0.0029$ & $0.015 \pm 0.0005$ & $0.109 \pm 0.0070$ & $0.696 \pm 0.0032$\\
        BT-VC-1 & $0.007 \pm 0.0012$ & $0.018 \pm 0.0025$ & $0.008 \pm 0.0005$ & $0.113 \pm 0.0063$ & $0.619 \pm 0.0044$\\
        \bottomrule
    \end{tabular}
    \label{tab:disentanglement-metrics}
\end{table}

However, abysmal gains along the compactness axes for higher dimensionality models like BT-2048 (MIG: 0.005) over BT-16 (MIG: 0.004) showcase the diminishing effect of dimensionality in contrast to BT-128 (MIG: 0.020) which potentially enable more granular decoupling. Furthermore, training with wider data variety as in BT-LS-360, improves both clustering and retention of compactness (MIG: 0.006, Modularity: 0.696), elucidating the value of diverse training corpora.

Therefore, while invariance and redundancy reduction induce minor factorization of informative factors of variation and disposal of irrelevant variation, additional explicit constraints must complement these objectives to achieve fine-grained disentanglement along speech factors like speaker traits, accents, emotions, and linguistic content throughout the latent feature hierarchy. Our analysis quantitatively demonstrates this limitation while revealing pathways to potentially facilitate targeted factorization through greater model capacity and diverse training data.

%
\subsection{Ablation of Loss Function Variants}
\begin{figure}[!h]
    \centering
    \includegraphics[width=0.45\textwidth]{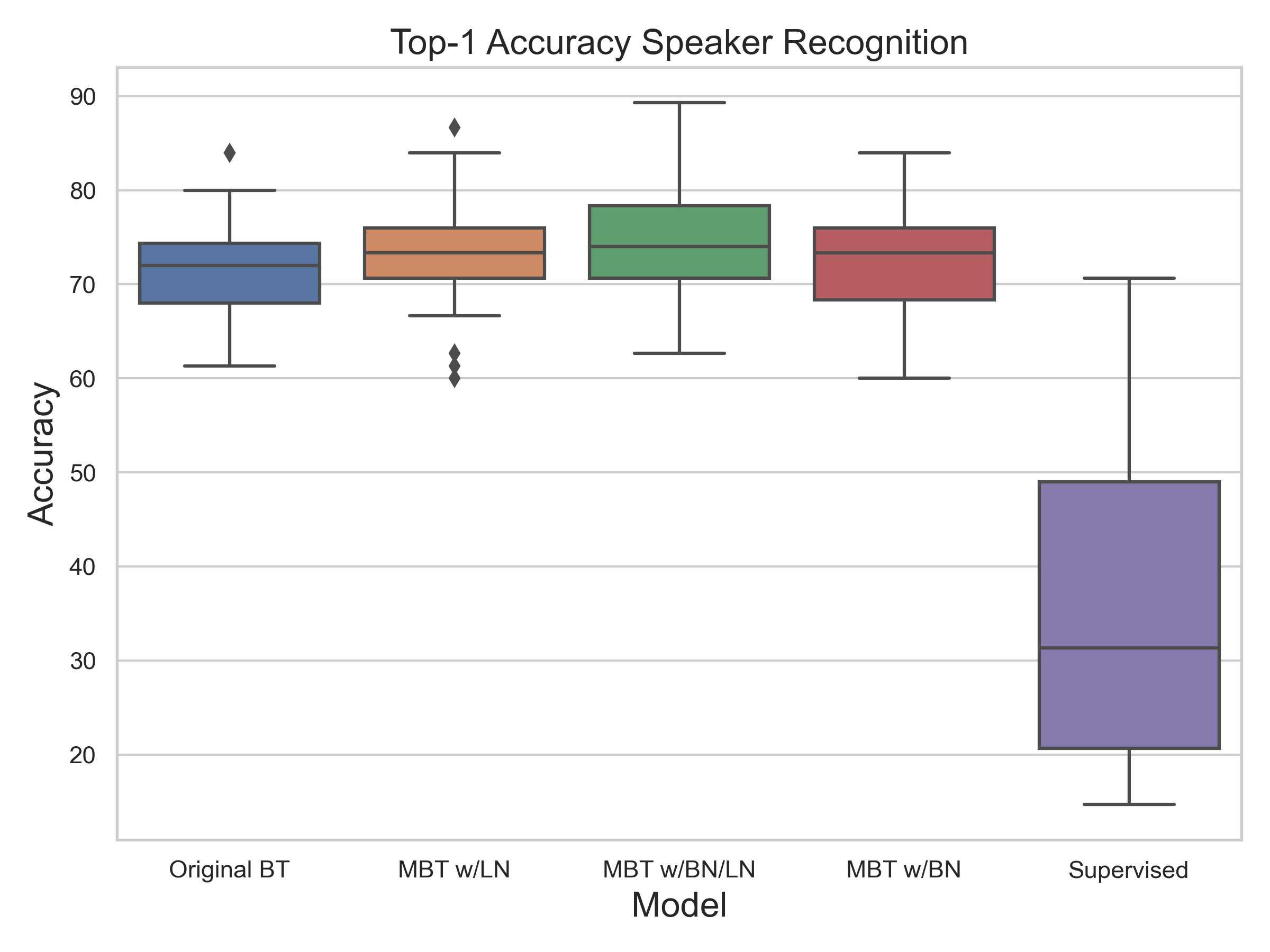}
        \includegraphics[width=0.45\textwidth]{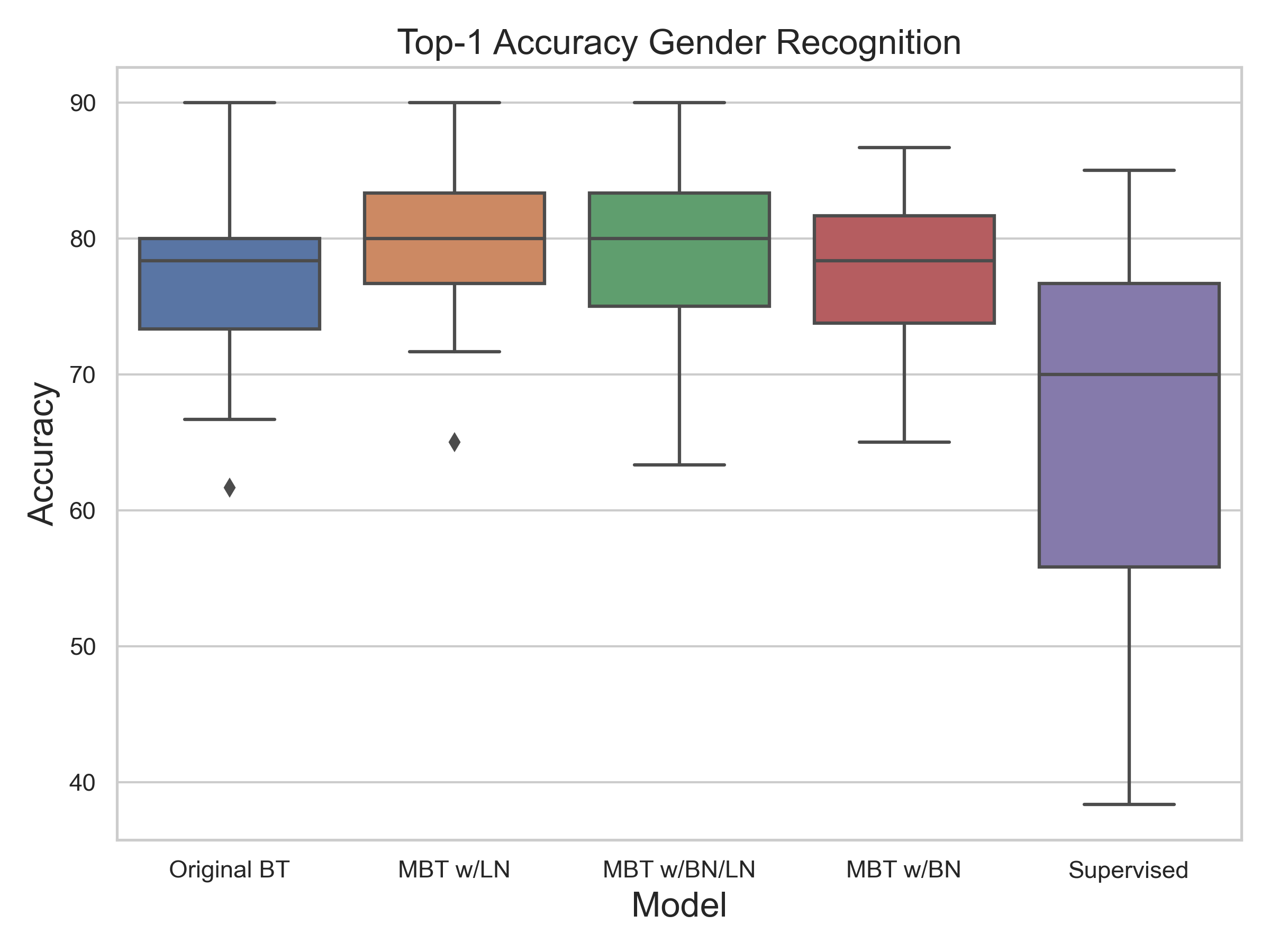}
        \caption{(a) Top-1 Accuracy for Speaker Recognition, comparing five base models over 50 experimental runs, highlighting the performance and stability of these techniques. (b) Top-1 Accuracy for Gender Recognition from speech, using the same base models, which shows a similar performance trend, indicating task-specific model effectiveness and the nuanced nature of gender features in speech data.}
    \label{fig:Speaker_Accuracy}
\end{figure}

\begin{figure}[!h]
    \centering
    \includegraphics[width=0.45\textwidth]{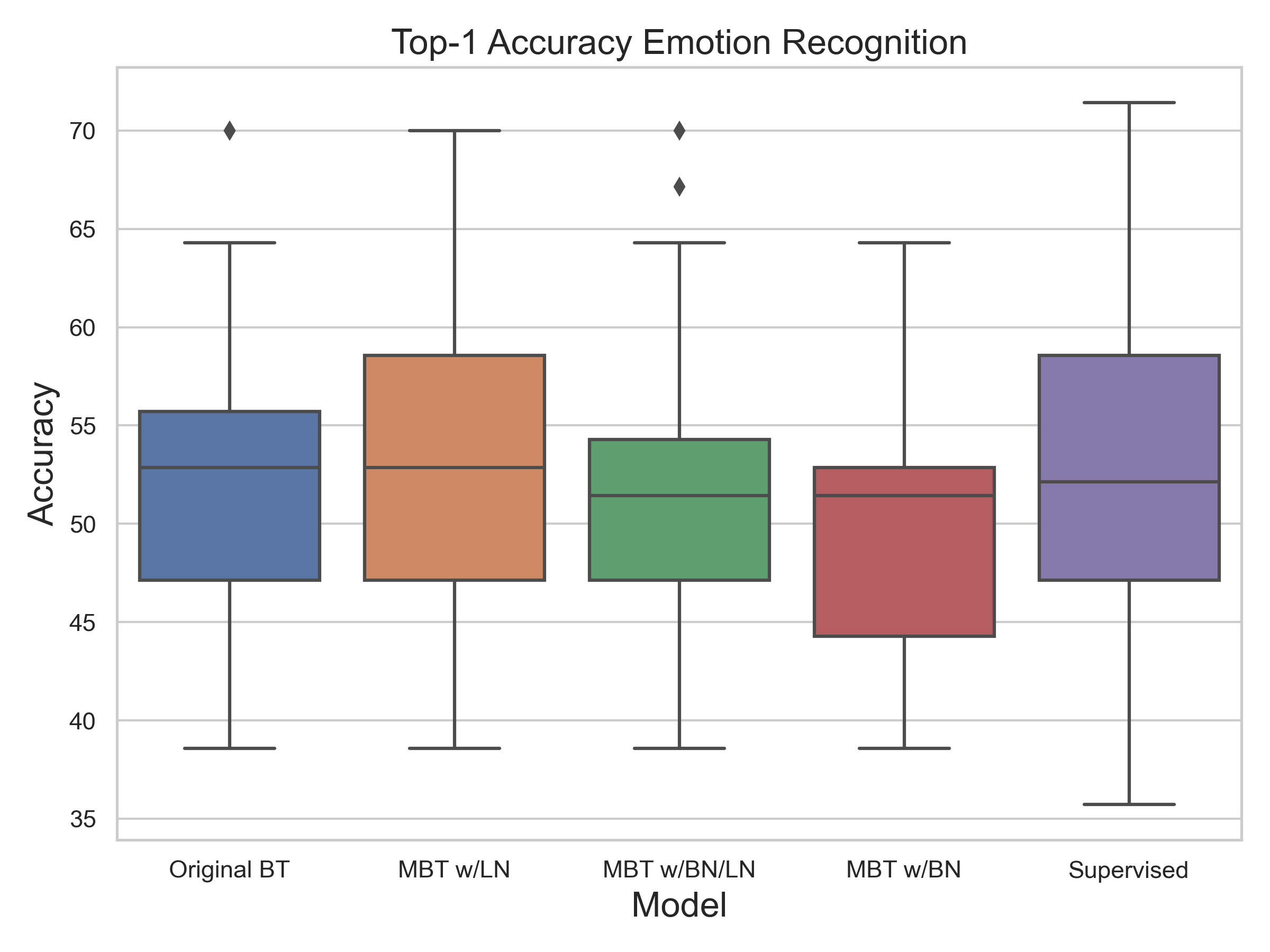}
        \includegraphics[width=0.45\textwidth]{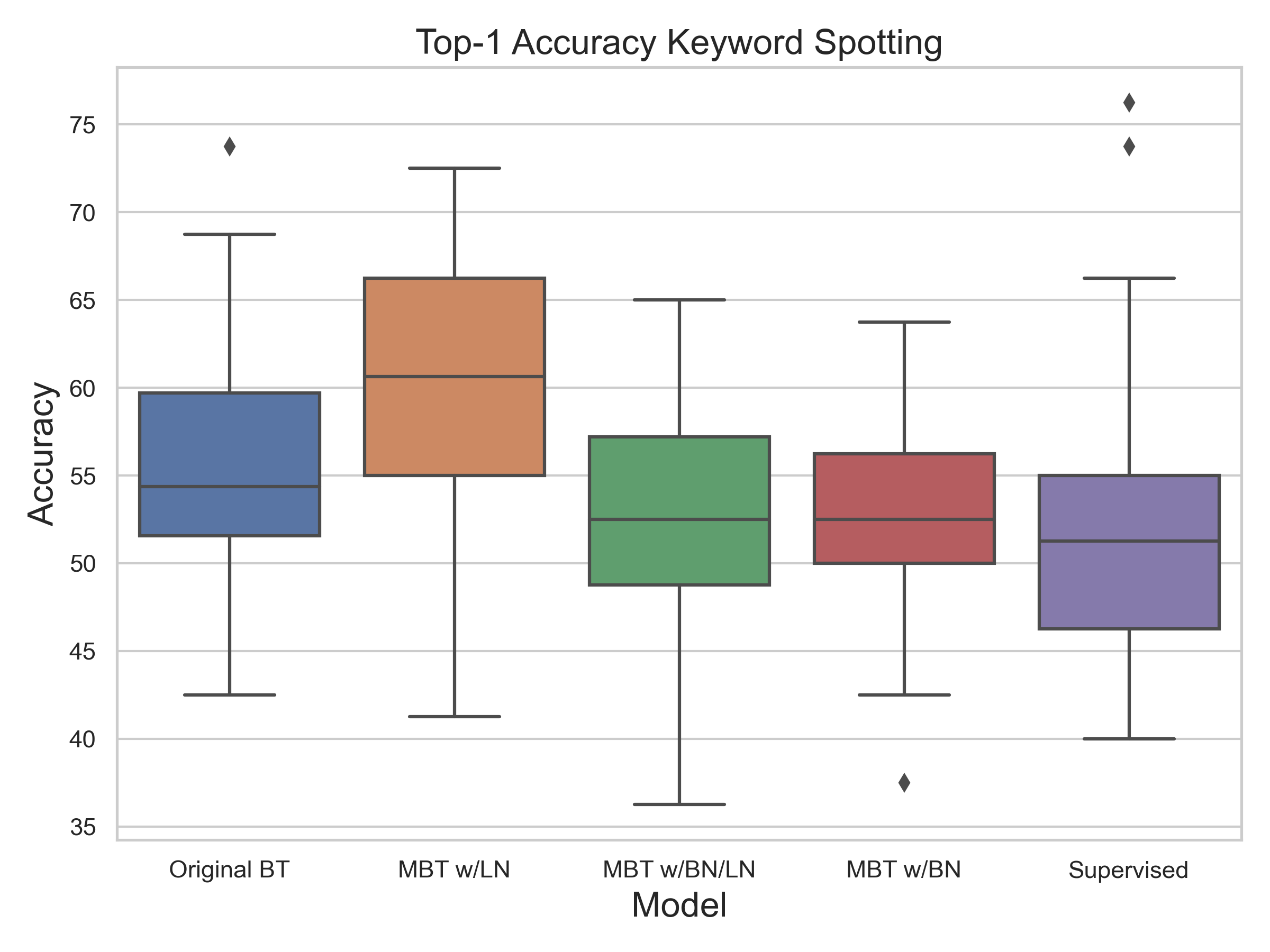}
   \caption{(a) Boxplot of Top-1 Accuracy in Emotion Recognition across five different base models over 50 experimental runs, showing the consistency and variability in model performances. (b) Boxplot of Top-1 Accuracy in a Keyword Spotting Task for the same base models and number of runs, illustrating the impact of model architecture on task-specific accuracy.}
    \label{fig:Emotion_Accuracy}
\end{figure}

In this section, we conduct an extensive ablation study on variants of the BT loss function, evaluating their impact on learned representation. Figures~\ref{fig:Speaker_Accuracy} and~\ref{fig:Emotion_Accuracy} present the results of this investigation, comparing the original Barlow Twins (BT) with several modified versions. Specifically, we analyze the Modified Barlow Twins with Latent Normalization (MBT w/LN), Modified Barlow Twins with Batch Normalization and Latent Normalization (MBT w/BN/LN) (column-wise), and Modified Barlow Twins with Batch Normalization (MBT w/BN). The ablation concludes with a benchmark using the standard Supervised method.

Building on this setup, the empirical results are illustrated in Figure~\ref{fig:Speaker_Accuracy} (left). This plot reveals the Top-1 accuracy of different models in the context of speaker recognition. We first note the Original BT model, which exhibits a median accuracy of around 70\%, paired with a relatively symmetrical interquartile range (IQR) and outliers that suggest variations in performance. In contrast, the MBT w/LN model demonstrates a similar median but with a notably tighter IQR, indicating more consistent results. A slight deviation is observed in the MBT w/BN/LN model, which has a marginally lower median accuracy and a larger IQR, pointing to more variability in its performance. A notable divergence is seen in the MBT w/BN model, characterized by a much wider range and lower median accuracy, with outliers indicating instances of particularly low performance. Interestingly, the Supervised model markedly stands out with a significantly lower median accuracy and a large IQR, underscoring its consistent underperformance relative to the other models. The presence of outliers, especially noticeable in the Original BT and MBT w/BN models, suggests instances where the models either excel or fall short dramatically. Overall, the MBT models incorporating layer normalization (LN) appear to strike a desirable balance between achieving high accuracy and ensuring result consistency, while the supervised model exhibits considerable limitations in accuracy for this specific task.

In Figure~\ref{fig:Speaker_Accuracy} (right), we can see the Top-1 accuracy of various models in gender recognition tasks. The Original BT model's median accuracy is situated just above 70\%, with a relatively broad IQR, indicating some variability in its performance. Notably, there are a few outliers that fall significantly below the lower quartile, which may point to specific instances where the model underperforms. Moving on to the MBT w/LN model, we notice a higher median accuracy and a narrower IQR, suggesting that this model not only performs better on average but also does so more consistently. The MBT w/BN/LN demonstrates a median accuracy comparable to MBT w/LN, but with a slightly wider IQR, indicating a bit more inconsistency in its results. In contrast, the MBT w/BN model exhibits a lower median accuracy and the widest IQR of all the MBT models, showing substantial variability in performance. Lastly, the Supervised model shows a significantly lower median accuracy, below 60\%, and a very wide IQR, which implies that while it can occasionally perform well, it is generally less reliable than the other models. The presence of outliers in the Original BT and MBT w/LN models suggests that there are occasional deviations in performance, which could be due to a variety of factors such as model overfitting, anomalies in the test data, or limitations inherent to the models themselves. Overall, the MBT w/BN/LN model seems to offer the best balance between accuracy and reliability for gender recognition tasks.

To extend our investigation beyond speaker representation, we further explore the impact of these loss function variants on emotion recognition and keyword spotting tasks. Figure~\ref{fig:Emotion_Accuracy} (left) showcases the results for emotion recognition accuracy, while the second plot highlights accuracy in keyword spotting. In this plot, we are comparing the Top-1 accuracy across different models for emotion recognition. The Original BT model shows a median accuracy slightly above 55\%, with a broad IQR which indicates a fair amount of variability in performance. The model also exhibits outliers, suggesting some predictions are notably different from the rest. The MBT w/LN model presents a higher median accuracy near 60\% and a slightly narrower IQR, implying more consistent performance than the Original BT. The MBT w/BN/LN has a similar median to the MBT w/LN but with an even tighter IQR, which may indicate a higher level of consistency in its emotion recognition capabilities. In contrast, the MBT w/BN shows a lower median accuracy and a wider IQR, indicating less reliability. Lastly, the 'Supervised' model shows a median accuracy comparable to MBT w/BN, but with the widest IQR of all the models, signifying the most variability in its accuracy. This analysis suggests that while no model excels at emotion recognition with high accuracy, the MBT w/LN and MBT w/BN/LN models perform more consistently than the others, with the supervised model being the least consistent and potentially overfitting or not generalizing well to the emotion recognition task.

Shifting our focus to the parallel task of keyword spotting, the subsequent part of Figure~\ref{fig:Emotion_Accuracy} (right) offers an intriguing comparison. Examining this, we observe the performance of various models in the task of Top-1 accuracy keyword spotting. The Original BT model has a median accuracy of just above 50\%, with a moderate IQR, suggesting a decent consistency in performance. However, there is a noticeable lower outlier that could indicate occasional significantly less deviation from the median. The MBT w/LN model presents a higher median accuracy, around 65\%, and a tighter IQR, which points to a more consistent and accurate performance in spotting keywords. MBT w/BN/LN shows a median accuracy comparable to MBT w/BN but with a slightly broader IQR, implying a bit more variability. The MBT w/BN model indicates a lower median accuracy, near 50\%, and a wider IQR, signifying less reliable performance. Finally, the Supervised model exhibits a median accuracy similar to MBT w/BN, but with the widest IQR of all the models, reflecting substantial inconsistency in its keyword spotting capability. Outliers in the Original BT, MBT w/BN/LN, and Supervised models suggest that certain keywords may be particularly challenging for these models. In summary, while all models show potential for keyword spotting, MBT w/LN demonstrates the best combination of high accuracy and consistency, with the supervised model appearing to be the least stable.

While the minor advantages of latent normalization were observed across the downstream tasks, our broader analysis of tasks paints a more nuanced picture. In this ablation study, we evaluated variants of the BT objective on several speech-processing tasks. Our results suggest that incorporating normalization into the loss can potentially improve model accuracy and reliability, depending on the specific downstream task. However, further investigation is needed to determine if these trends hold more broadly, as the benefits were not conclusively demonstrated for all tasks. While variants like Modified Barlow Twins with Latent Normalization showed promise, claiming definitive improvements would require more extensive experimentation and analysis. This study provides an initial path that modifying the Barlow Twins objective may yield benefits, motivating further research into enhanced SSL techniques.
\section{Discussion}
\label{dis}
This study provides an extensive empirical analysis of the Barlow Twins (BT) framework for SSL in speech representation. Our findings affirm the framework's effectiveness, while also highlighting critical areas for further development and exploration.

\subsection{Generalization in Downstream Tasks}
The BT framework demonstrates notable success in generalization across various downstream tasks, such as speaker recognition, gender detection, emotion recognition, and keyword spotting. Remarkably, models trained on LibriSpeech-100 achieved over 80\% accuracy in speaker identification with only half of the labeled data, suggesting that the curated quality of a dataset may be more crucial than its size. This insight opens up opportunities for optimizing dataset selection in speech processing tasks, focusing on quality and diversity rather than volume alone.

However, the transferability of learned representations is influenced by task complexity and domain alignment. Simpler tasks, such as speaker recognition, benefit more rapidly from pre-trained models, while more complex tasks like emotion detection necessitate greater amounts of domain-specific data. This variation underscores the need for task-specific fine-tuning and adaptation of pre-trained models, particularly when dealing with complex or nuanced speech-processing tasks.

\subsection{Disentanglement of Latent Representations}
Our disentanglement analysis reveals a significant area for improvement in the BT framework. Despite achieving redundancy reduction and invariance, the framework falls short in optimally disentangling key explanatory factors in speech, as indicated by the low MIG and SAP scores. This limitation points to the necessity of integrating additional mechanisms or constraints to enhance the disentanglement capabilities of the framework. The potential for leveraging greater model capacity, as shown by the improved compactness in higher-dimensional models, and the benefits of diverse training data, suggest paths forward. Future research should focus on developing and incorporating novel architectural models and inductive priors that can facilitate more effective and targeted factorization of speech attributes.

\subsection{Inconsistencies Across Different Tasks}
The ablation studies conducted as part of this research provide valuable insights but also highlight inconsistencies across different tasks. While improvements were observed in certain scenarios, such as emotion recognition and keyword spotting with latent space normalization, these were not uniformly seen across all tasks. This inconsistency calls for a more nuanced understanding of how different components of the loss function and other architectural choices affect various speech-processing tasks. Further investigation and experimentation are needed to establish more definitive conclusions about the efficacy of these modifications.



\section{Conclusion}
\label{con}
This work provides a thorough empirical evaluation of the Barlow Twins framework for self-supervised speech representation learning. Our findings validate the efficacy of this approach in achieving generalization across diverse downstream tasks, underscoring the importance of dataset quality over size. However, results also reveal limitations in disentangling explanatory factors within the learned representations, despite redundancy and invariance constraints. Additional techniques are needed to enable fine-grained factorization of key speech explanatory factors. Ablation studies isolate gains attributable to invariance but inconsistent advantages motivate enhancements to the framework. To address these limitations and advance Barlow Twins for speech, incorporating perceptual principles, speech-specific pretext tasks, and comparative benchmarking are proposed. In summary, this investigation substantiates the sample efficiency and emerging utility of Barlow Twins while paving the path for continued progress through our rigorous assessment methodology. Key challenges exist in realizing fully decoupled hierarchical representations, motivating tailored pretext tasks and constraints to enable more granular speech attribute factorization. Overall, this work not only validates the potential of self-supervised learning for speech processing but also opens avenues for enhancing these techniques toward more efficient, and robust representations.


\vspace{6pt} 




\authorcontributions{``Conceptualization, S.P. and Y.B.; methodology, Y.B.; software, Y.B.; validation, U.K. and Y.B.; formal analysis, Y.B. and U.K.; investigation, Y.B.; resources, Y.B.; data curation, Y.B.; writing---original draft preparation, Y.B.; writing---review and editing, Y.B. and U.K.; visualization, Y.B.; supervision, G.H. and S.P.; project administration, Y.B.; funding acquisition, S.P. All authors have read and agreed to the published version of the manuscript.''}

\funding{
This work was funds of the research training group (RTG) in ``Computational Cognition'' (GRK2340) provided by the Deutsche Forschungsgemeinschaft (DFG), Germany, and an EU-Consolidator grant (772000, TurnTaking).
}

\institutionalreview{Not applicable}

\informedconsent{Not applicable}

\dataavailability{All data are publicly available and  accessible.}



\conflictsofinterest{The authors declare no conflict of interest.} 




\appendixtitles{no} 
\appendixstart
\appendix



\begin{adjustwidth}{-\extralength}{0cm}

\reftitle{References}


\bibliography{references}

\PublishersNote{}
\end{adjustwidth}
\end{document}